# SCHOOL MANAGEMENT INFORMATION SYSTEMS: CHALLENGES TO EDUCATIONAL DECISION-MAKING IN THE BIG DATA ERA

Vivienne V. Forrester

College of Engineering and Computing, Nova Southeastern University, Fort Lauderdale, Florida, United States of America

*ABSTRACT*

*Despite the benefits of school management information systems (SMIS), the concept of data-driven school culture failed to materialize for many educational institutions. Challenges posed by the quality of data in the big data era have prevented many schools from realizing the real potential of the SMIS. The paper analyses the uses, features, and inhibiting factors of SMIS. The paper proposes a five-phase conceptual model that assist administrators with making timely, quality decisions. The paper enriches the theoretical landscape of SMIS usage in the era of big data and lays a foundation for the future by establishing an educational decision-making model.*

*KEYWORDS*

*School management information system, big data, educational decision-making, data-driven schools, educational model, student-information system.*

## 1. INTRODUCTION

Developments in information technologies and other factors such as information exchange, increasing expectations of the society, current managing perceptions and applications have made a significant impact on organizations and their practices. With the introduction of the 2001 No Child Left Behind (NCLB) Legislation, the use of data to inform educational decisions has garnered increased attention, primarily spurred by the need for accountability of gathering, aggregating, and reporting of student-level data at the state and federal levels of education. In recent years, school districts have invested millions of dollars in integrating and improving the computer systems schools use to store student data [1]. The isomorphic shift from individualized decision-making about education to a large-scale data-driven context has changed the nature of national educational decision [2].

The school administrative computer applications development, started in the late 1970s, during this phrase, the use of information technologies in educational institutions was used mainly to improve the efficiency of school offices, such as store student and staff data [3]. This was a significant achievement for institutions; however, despite increased access to data systems, systems often lack the kinds of data that educators find most useful for instructional decision making [4]. The use of School Management Information Systems (SMIS) was one form of information technology adopted to fill the gap, due to its ability to generate data and its efficiency and effectiveness to save time and facilitate the development of alternative solutions for sophisticated problems.

   

The emergence of big data attracts the attention of many industries including academia, as the amount of global data expands exponentially, with the data unit moving from GB and TB to PB [5]. Obtaining and evaluating big data from various sources and with multiple uses provide a substantial benefit for educators and administrators for understanding students' needs, improving teaching methodology, courses and curriculum, and predicting students' attendance and academic performances. However, schools must base educational decision-making on accurate, timely, and high-quality data, coupled with the relevant skilled personnel. Therefore, it is essential to understand the challenges of SMIS in the era of big data.

Laudon and Laudon [6] described an information system "as a set of interrelated components that collect (or retrieve), process, store, and distribute information to support decision-making and control in an organization." A SMIS also referred to as student information system, education information system, student management systems, academic administration system or student administration system is a "management information system that is designed to match the structure, management task, instructional processes and special needs of the school" [7].

## 1.1 Specific Problem

According to Hinze-Pifer and Ramsey in [1], as computers became an integral part of schools at the end of the 1990s, schools began to digitize student data and use databases to organize student records, from standardized test scores to health records to demographic information. However, despite the widespread use of databases in United States schools, they are merely information storage systems, without making applicable information readily available to essential users. Additionally, as the concept of big data increases, school administrators and educators are faced with the challenge of sorting an enormous amount of data in SMIS in a limited time-frame. Hinze-Pifer and Ramsey in [1] further argued that many school districts use multiple databases such as one for attendance, one for grades and one for test scores that rarely can communicate with each other.

Additionally, New [8] pointed out that despite most U.S. schools are integrating information technology (IT), with 93 percent of teachers frequently using technological devices in their lessons, most schools still focus on using IT to sustain operations, rather than leverage data to transform and enhance these operations. As a result, U. S. schools and educators are mainly failing to use data to make timely decisions for students and to transform the education system [8]. The strengths and drawbacks of handling data in schools have received attention from both the academic and popular press [9], but research is still lacking. Shah [10] made a call for studies on SMIS to concentrate on the methods of improving the use of SMIS by school administrators.

## 1.2 Goal of the Study

The main goal of this study is to present a conceptual model that will guide administrators to identify and eliminate the inhibiting factors that will impact their ability to use school management information systems effectively. The conceptual model will also aid administrators with identifying key variables that will influence the inhibiting factors, towards effectively managing decision making. Following the descriptive decision-making theory, the model will present administrators with a set of principles that will enhance the use of information systems in schools. Additionally, the paper will guide future researchers to understand the features and uses of school management systems, as well as the limitations and how to address them, to make SMISs more useful and practical for educational management.

## 2. REVIEW OF THE LITERATURE

A data-driven education system primed for effective decision-making will encompass a variety of data-focused educational technologies including school management information system, learning management systems, and Data warehouses [8]. In the big data era, policymakers must also be concerned about the quality of data in the SMIS. Big data introduces new features; hence its data quality also faces many challenges. The characteristics of big data include the 4Vs: Volume, Velocity, Variety, and Value [11]. Volume refers to the enormous quantity of the data, measured in TB or above. Velocity refers to the extraordinary speed at which data are being formed and its immediate need for data analysis. Variety indicates that there is a diversity in the kinds of data types including structured data and unstructured data, requiring higher data processing capabilities and value represents the relevance of the data mined from the system, and the ability of policymakers to convert enormous amounts of data into actionable results [5]. School administrators can improve their decision making by leveraging big data with effective SMIS implementation.

### 2.1. School Management Information Systems

School Management Information Systems are digital tools that are being used in schools to support a variety of administrative tasks such as monitoring attendance, assessment records, reporting, financial management, and resource and staff allocation [10]. According to Vischer (1996b), "information systems in schools can provide administrators with the information required for informed planning, policy-making, and evaluation." Gurr [12] claimed that such systems had transform school management in the areas of communication, governance, and decision making.

In the early 2000s principals initiated the use of information systems increasingly to manage daily tasks. Generally, to increase managerial effectiveness by processing information, and gaining superiority in competitions by directing strategies to support ongoing student assessment, especially by facilitating formative and summative evaluations [13] to manage changes in the instruction process, the education environment, and determining the needs of students [12]. In many cases, SMIS are used to simplify or automate routine classroom administrative tasks such as recording attendance [14] and maintain communication with parents [15]. Strategically, SMIS help managers to determine the aims of schools, make long term plans, distribute resources, form educational methods of the future, and assess performances of teachers and success of the school [7]. Schools use SMISs to make decisions such as student placement in courses and programs at the middle and high school levels [4].

School management information systems are also useful to digitize data, facilitate communication between teachers, administrators, parents, and other stakeholders, and provide easy access to data about student performance, thereby increasing administrative efficiency and reducing educator's workload [10]). According to Gallagher, Means, and Padilla [16] having a more detailed understanding of each student's progress, educators can hopefully better discuss the information with parents, and teachers can also discuss student progress and instructional strategies among each other and with administrators [17]. Kaya and Azaltun [18] found that communication and information sharing among stakeholders provides for more effective decisions. SMIS can also inform educators about their students' non-cognitive skills, relating to grit, motivation, self-efficacy, and resilience. These soft skills may not unequivocally relate to academic proficiency or grades but are still critical to students' overall performance.

To this end, SMIS should enhance the administrator's work functions with a reduced workload, accurate, up-to-date and timely decisions to manage institutions efficiently and effectively. Decision making is at the heart of educational management, due to the complex and unexpected nature of the school environment and increasing expectations from educational systems. As a result, school administrators must gather and analyze information continuously, make decisions in shorter times, more frequent and for more complicated schools [19]. As education software corporations generate new analytical strategies and as the big data era surges, the value of SMIS will be more impactful for educational decision-making. This concept was evident in the Tacoma, Washington public school district that integrated predictive analysis with the data from its SMIS to build intermediation strategies that boosted its high school graduation rate by 27.6 percent to 82.6 percent, over six years (Tacoma Public Schools, 2016).

## 2.2. Features of School Management Information Systems

Typically, SMIS accumulate data into web-based data portals for easy access to the user. These data can be tailored to meet the needs of students, educators, parents, and other stakeholders [15]. Prior research suggests SMIS should have a variety of specific features to be successful. They should compile and allow for the analysis of multiple data for relevant stakeholders, including teachers, administrators, and parents. Data elements such as student demographics, attendance, formative and summative assessment results, standardized test scores, special programs, and English language learning (ELL) evaluations results should be included [4].

The system should be capable of mining and generating specified reports, including detailed queries and ad-hoc reports, from available data to enhance decision making, as well as capable of assisting with curriculum evaluation and teacher planning through Built-in benchmark assessment tools [20]. Generally, the features of systems should include elements that support teacher and administrator decision making, in a format that is relevant and easily accessible.

New [8] posited that data-driven institutions SMIS should support four critical goals. The systems should support personalization, focus on evidence-based learning, aim to improve school efficiency and foster continuous innovation. According to New [8] personalization is education that is "customized, dynamic and tailored to the abilities, interests, and needs of a particular student based on that student's data and informed by historical data." Evidence-based learning refers to educators having quick and easy access to the record, the ability to evaluate, distribute, and utilize data to inform every aspect of their decision-making, as they learn from new information, and expand and disseminate effective strategies, through the community of practice. School efficiency focuses on the application of big data to analyze the correlation between educators and administrators practices and student and school performance. Continuous innovation aims at the ability of SMIS to provide relevant data to specified stakeholders in a form that is valuable for improved decision-making, discover new insights, and generate predictive models.

## 2.3. Inhibiting factors of School Information Management Systems

Despite the vast availability of desired features, functions, and valuable uses for school management information systems, the literature suggests that several factors inhibit their current use in American schools [4]. As highlighted in [21] with the availability of an enormous of amount of data, many schools have been "data rich," however, they were also "information poor" as the data access was not user-friendly to most policymakers.

the According to Shah [10], several challenges to SMIS use in educational management are apparent in the literature, including low self-efficacy, lack of timely information, limited coaching and preparation, the absence of executive support, and the paucity of technical assistance. New [8] supported this concept by highlighted several major inhibitors in SMIS data for effective decision-making in U.S. schools including inadequate teacher training, flawed data infrastructure, and privacy fears.

Schifter et al. [22] indicated that the use of data from school management information systems in decision making for schools is no longer a choice. However, underlying issues continue to plague the process and its impact on practice for a variety of reasons, including timely availability of data, accessibility to data, and administrator understanding of how to effectively use the data generated. Timeliness of data is an inhibitor as the results of state and national assessments are often accessible after students have advanced to the subsequent grade, making the data less useful [23]. School management information systems should make data relevant; in doing so, administrators should ensure that the data is timely and sensitive to the curriculum of each teacher [23]. Inadequate training, professional development, or support systems also hinder the use of SMIS [20], [23].

As pointed out in [24] user attitudes also inhibit the success of school management information systems. Users' concerns about the linkage of the system to accountability policies, testing systems, and the processes of the curriculum may lead to opposition, as teachers often feel reliance on evidence-based practices reduces their autonomy, or that data will be used to punish them. In other instances, educators lack the relevant skills to effectively leverage data. Mandinach et al. [25] supported this concept by highlighting that most staff lack training on how to turn data into relevant information. According to the center for data innovation in 2014, as a prerequisite for teacher licensure, only 19 states require educators to demonstrate data literacy skills, including the aptitude for assessing the accuracy of a dataset and the expertise to transform data into actionable results.

## 3. APPROACH

### 3.1. Conceptual Model

The concept model, shown in Figure 1, is centered on the descriptive decision theory [26]. Decision theory was developed since the middle of the 20th century as an interdisciplinary approach that theorizes about human make decisions. As human's behavior is often entrenched in making decisions; to theorize about the decision is primarily to theorize about human activities. More specifically, it focusses on how humans use freedom to choose between options for goal-directed behavior. "Classical theories of choice in organizations emphasize decision making as the making of rational choices by expectations about the consequences of an action for prior objectives, and organizational forms as instruments for making those choices" [27]. Most organizations would like to think employees follow such rational processes in decision making, however, in practice, it is infrequently the case, as there is a gap between normative theory (what people should do) and descriptive theory (what people do or have done) [28].

As administrators receive data from school management information systems, it is essential to take into consideration the inhibiting factors that may affect the quality of such data and the system in general. It is also very critical to examine variables and constructs of those inhibiting factors that when eliminated from the process will enhance the use of school management information systems by administrators and educators and assist them in making better quality decisions. The proposed model shown in figure 1 illustrates the relationship among the inhibiting factors, the design constructs and the descriptive theory principles that may result in the improved use of data.

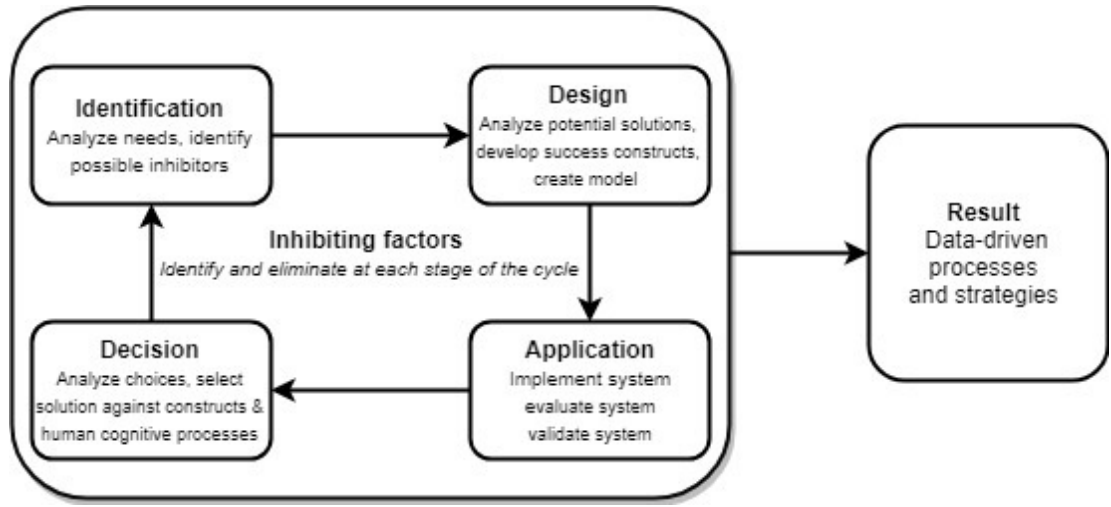

Figure 1. School management information systems educational decision-making model

This paper developed a SMIS educational decision-making model to highlight fundamental considerations of five phases for data-driven schools: identification, design, application, decision, and result. Inhibiting factors are identified, evaluated, and eliminated, as necessary at each phase of the model.

**3.1.1. Identification Phase**

Many SMISs are implemented based on assumed benefits, which are often poorly specified, which can result in implementation and usage challenges. Often terms such as "improved student performance" and "improved efficiency" are used, detailed results from specific system functionality are difficult to measure and to predict, as most implementations require major modifications to operational practices [31]. Thus, many schools often grapple with the challenges of conceptualizing the necessary short, medium, and long-term transformations. In the identification phase, schools should conduct a thorough mapping of the existing operational practices, and establish a needs analysis of the school, including the needs of all stakeholders. In this phase, the groundwork would result in agreement on the type of data that the system should store, and all possible inhibiting factors such as lack of infrastructure, lack of training, and technical support. School should identify possible strategies to eliminate the inhibiting factors. For instance, the type of training that will be executed to support educator and administrators. To overcome user resistance, school districts and leaders must seek "buy-in" from teachers and administrators early in the process, often in the identification phase, and establish strategies to reduce or eliminate fears of using the system.

**3.1.2. Design Phase**

In the design phase, schools should focus on the needs of all stakeholders including administrators, educators, support staff, and parents, according to a strategic vision of the end goal of the SMIS for decision-making. This process may involve aiming for school-wide changes or to concentrate on streamlining specific processes such as attendance and record-keeping. Extant literature from the field of organizational change has argued that high-level leaders are crucial for technology implementation and adoption. However, it is also essential to involve buy-in of different professional stakeholders [31], such as teachers, counselors, librarian, and administrative staff. In the design phase, the school evaluates if a SMIS should be designed or adapted according to the needs of the schools and building on the identification phase, establish concrete efforts to ascertain further buy-in from stakeholders. It is also crucial to anticipate educators and administrators' data needs to be delivered in real-time and consider the speed of data accessibility, delivery methods, and the potential quality of the data.

Developers and schools should highly consider information systems success constructs and usability features in the design phase. Usability principles and success constructs will augment the transition of the system into the educator's workflow. Nielsen [32] usability heuristics indicate that systems should be designed to deal with the requirements of the organization and its users, an interface should be clear and easy to navigate, should provide accurate and up-to-date information, should generate reports that are easily customizable and include built-in benchmark and other assessments tools. Developers must make it easy for educators to "do the right thing, consistently." At the design phase, strategies to eliminate possible inhibiting factors should also be implemented, such as the acquisition of appropriate technology infrastructure for the smooth implementation of the system.

### 3.1.3. Application Phase

In the application phase, the system is implemented and evaluated against usability principles. Schools may adopt an evaluation system to evaluate the core elements of the SMIS. Whyte and Bytheway [33] framework for assessing an information system effectiveness focuses on the product, that is, hardware, software, and training provided to users; service, that is, how users receive feedback; and the process by which the product and service are provided. Further inhibiting factors are also identified, addressed and eliminated if possible. After implementation, administrators and educators should engage in training and professional development. Administrators and educators should participate in training on how to use the system, and how to use the information generated by the system for decision making at the initial stage of system implementation as well as on an ongoing basis [34].

### 3.1.4. Decision Phase

According to Bates et al. [29], "speed is everything." The researchers indicated that the velocity of an information system is the factor that users attribute the most value. With the onset of big data, data in an enormous quantity is available at relatively fast speed. However, a SMIS should present the data in a format that is useful for administrators and educators at the appropriate level and time. Following the descriptive theory approach, humans usually do not behave in ways consistent with axiomatic rules, identifies the best decisions to be made, fully informed, capable of computing with perfect accuracy or act entirely rational. Instead, they rely on their knowledge, values, and beliefs. In the decision phase, policymakers should recognize the nature by which humans make decisions and take into considerations the factors that may affect their decisions. Therefore, by heightening the success design constructs and reducing the inhibiting factors, administrators will be more equipped to analyze choices, select from options and make timely, quality decisions, leading to productive results.

International Journal on Integrating Technology in Education (IJITE) Vol.8, No.1, March 2019

**3.1.5. Result Phase**

As highlighted in [35], using technology in schools is progressive. Managing the knowledge within the information system and controlling other factors such as consciously training staff in useful data analysis will be essential to the continued generation of data-driven processes and strategies within educational institutions. As posited by Creswell et al., [31], real-time longitudinal data collection strategies, while generating formative feedback are necessary for the maintenance of systems, as information output from parts of the system will become data input for other aspects. Understanding this input-process-output cycle is essential to the overall educational decision-making process of SMIS, as results generated as "bad" output may contribute to ineffective decisions.

## 4. CONCLUSION

The literature indicates that a school management information system can be a useful tool in educational management. Schools use SMISs in various ways to support and enhance administrative decision making. However, there are various inhibiting factors including lack of infrastructure, lack of training and support, and administrator's inability to effectively analyze data that directly impacts the use of the systems in schools. As policymakers continue to promote the implementation and application of SMIS, they should be cognizant of the possible constraints of the system as well as the unique requirements of different users. Schools should implement strategies to eliminate inhibitors while increasing success approaches based on usability constructs principles. Adoption of the SMIS educational decision-making model can assist in the identification and elimination of the challenges associated with SMIS for decision-making in the big data era. Implementing the SMIS educational decision-making model can further guide schools to fully transition the American K-12 education into the 21st data market. Schools can rely on data to make improved operational and strategic decisions and innovate new solutions to educational problems.

Future research should examine the role of universities and other higher institutions in the training of administrators and educators in the use of SMIS in education and leadership programs for educational decision-making. Studies on SMIS should also examine the impact of the integration of SMIS on users' satisfaction and its acceptance in educational leadership based on the SMIS model for decision-making.

## AUTHOR

Vivienne Forrester is an educator, instructional technologist, and a technology integration coach. With over fifteen years in the education industry, she is passionate about STEM education and profoundly believes in life-long learning. Vivienne Forrester currently serves as the head of upper school academic technology and computer science teacher at The Chapin School, New York. Prior to joining Chapin, Vivienne Forrester served as an International Baccalaureate (IB) information technology teacher, technology department chair, Project Lead the Way (PLTW) coordinator, and STEM coordinator with the public charter school system in Washington, DC. She also worked for the National Training Agency in Jamaica as a computer science instructor. She is a graduate of the University of Technology, Jamaica and Nova Southeastern University (NSU) with majors in Computing with Accounting (B.Ed) and Management Information Systems (MSc). Ms. Forrester is a Ph.D. candidate in Information Systems at NSU.